\documentclass[floats,aps,twocolumn,superscriptaddress,showpacs]{revtex4-1}
\usepackage{amsmath}
\usepackage{epsfig}
\usepackage{color}
\usepackage{endnotes}
\usepackage{hyperref}
\usepackage[normalem]{ulem}
\newcommand{\be}{\begin{equation}}
\newcommand{\ee}{\end{equation}}

\newcommand{\ber}{\begin{eqnarray}}
\newcommand{\eer}{\end{eqnarray}}

\newcommand{\ket}{\rangle}

\usepackage{changes}

\begin{document}

\title {First principles description of the giant dipole resonance in $^{16}$O}

\author {S. Bacca}
\affiliation{TRIUMF, 4004 Wesbrook Mall, Vancouver, BC, V6T 2A3, Canada }

\author{N. Barnea}
\affiliation{Racah Institute of Physics, Hebrew University, 91904, Jerusalem} 

\author{G. Hagen} 
\affiliation{Physics Division, Oak Ridge National
  Laboratory, Oak Ridge, TN 37831, USA} \affiliation{Department of
  Physics and Astronomy, University of Tennessee, Knoxville, TN 37996,
  USA} 

\author{G. Orlandini} \affiliation{Dipartimento di Fisica,
  Universit\`a di Trento and Istituto Nazionale di Fisica Nucleare,
  Gruppo Collegato di Trento, I-38123 Trento, Italy }

\author{T. Papenbrock}
\affiliation{Department of Physics and Astronomy, University of
  Tennessee, Knoxville, TN 37996, USA} \affiliation{Physics Division,
  Oak Ridge National Laboratory, Oak Ridge, TN 37831, USA}

\date{\today}

\begin{abstract}
  We present a calculation of the giant dipole
  resonance in $^{16}$O based on a nucleon-nucleon (NN) interaction from
  chiral effective field theory  that reproduces NN scattering data with 
  high accuracy.
  By merging the Lorentz integral
  transform and the coupled-cluster methods, we extend the previous
  theoretical limits for break-up observables in light nuclei with
  mass numbers ($A\le7$), and address the collective giant dipole
  resonance of $^{16}$O.  We successfully benchmark the new approach
  against virtually exact results from the hyperspherical harmonics
  method in $^4$He. Our results for $^{16}$O reproduce the position
  and the total strength (bremsstrahlung sum rule) of the dipole 
  response very well.  When
  compared to the cross section from photo-absorption experiments the
  theoretical curve exhibits a smeared form of the peak. The tail
  region between 40 and 100~MeV is reproduced within uncertainties.
\end{abstract}


\maketitle

{\it Introduction.}--- Giant resonances dominate the dipole response
in a variety of self-bound quantum systems such as 
nuclei ~\cite{Berman1975}, atomic clusters~\cite{brechignac1989}, and
fullerenes~\cite{bertsch1991}.  In photonuclear cross sections, the
giant resonance is a very pronounced peak at about
10--20~MeV.  After its first observation in Uranium by Baldwin and
Klaiber in 1947~\cite{BaK47}, this prominent structure has been
observed across the nuclear table. The first theoretical
interpretations were given by Goldhaber and Teller~\cite{GoT48}, who
proposed that the peak had to be of dipole nature [hence the
name giant dipole resonance (GDR)], and by Steinwedel and
Jensen~\cite{steinwedel1950}. The resonance was ascribed
to the collective motion of protons against neutrons, and its width
to the transfer of energy from this vibration into other modes
of the nuclear motion~\cite{Speth1981, Bertsch1983}. Since then the GDR has been
the object of numerous theoretical studies trying to account
for the GDR centroid and width in collective as well as microscopic
many-body approaches. Self-consistent mean field theories have been extensively applied to the description of GDR (see e.g.~\cite{Erler2011, Nakatsukasa2012} for recent reviews). The effect of the continuum (pioneered in~\cite{Shlomo}) and of the particle-phonon coupling (see e.g.~\cite{Colo2001,LyT12}) have been studied. However, in these calculations the dynamics is described by effective interactions which are fitted to medium-mass nuclei.

First principles continuum computations of the GDR, where one starts from interactions that reproduce NN phase-shifts, are only available for nuclei with mass numbers $A\le 7$. The reason is
that one not only needs to account for
complicated ground-state correlations, but in principle also the
many-body scattering problem must be solved.  In  few-body
calculations these challenges were overcome 
by employing the Lorentz integral transform
(LIT)~\cite{efros1994,Efl07}. This method reduces the continuum
problem to a bound state-like problem and thereby overcomes one of the
main challenges. The few-body bound state problem was solved with the
hyperspherical harmonics (HH) expansion~\cite{barnea2001,fenin1972}
and the no-core shell model~\cite{ncsm}. Starting from
semi-realistic NN interactions, this approach
computed the ``resonant shape'' of the photonuclear cross section of
$^4$He~\cite{EfL97, stetcu2007}, $^6$Li/$^6$He~\cite{bacca2002,BaB04} and
$^7$Li~\cite{BaA04}. For $^4$He, modern realistic two- and three-body
forces were also employed~\cite{gazit2006,Qua07}.
 
The studies of the $A=6$ isotopes~\cite{bacca2002,BaB04} showed that
the halo structure of the rare isotope $^6$He differs considerably
from the stable $^6$Li nucleus. In $^6$He, the dipole cross section exhibits two
well separated peaks, while a single resonant shape is observed for $^6$Li. These
results do not depend on the employed NN interaction. It is
interesting to extend such studies to heavier exotic nuclei
such as the  neutron-rich
isotopes of Oxygen and Calcium. The dipole cross sections of exotic Oxygen
isotopes show fragmented low-lying strengths and thereby differ
from the stable $^{16}$O~\cite{sagawa1999,Lei01}.  The first step
towards this goal is the extension of the
{\it ab-initio} LIT method  to heavier systems.
In this Letter, we  combine the LIT with the coupled-cluster (CC)
method~\cite{kuemmel1978,bartlett2007}, and microscopically compute
the GDR for $^{16}$O.  Our calculation employ  NN interactions from
chiral effective field theory~\cite{epelbaum2009,machleidt2011},
where no parameter is fit to nuclear data with $A>2$.


{\it Method.}--- The inclusive response function, also called dynamical
structure function, to a dipole excitation, is defined as
\begin{equation} \label{response}
  S(\omega)=\frac{1}{2J_0+1}\sum_f |\langle \psi_0
  |{\hat{D}_0}|\psi_f\rangle|^2
\delta (E_f -E_0 -\omega )\:\mbox{.}
\end{equation}
Here $\omega$ is the excitation energy, $J_0$ is the nuclear
ground-state spin, while $\psi_0$ and $\psi_f$ denote the ground 
and final state wave functions with energies $E_0$ and
$E_f$, respectively.  The translationally invariant dipole operator is
\be 
\hat{\bf D}=\sum_i^A
P_i \left({\bf r}_i - {\bf R}_{\rm cm} \right) = \sum_i^A
\left(P_i - \frac{Z}{A}\right) {\bf r}_i \,. 
\ee 
Here ${\bf r}_i$ and ${\bf R}_{\rm cm}$ are the coordinates of the $i$-th particle and
the center-of-mass, respectively, while $P_i$ is the
proton projection operator and $Z$ is the number of protons. 
In Eq.~(\ref{response}) $\hat{D}_0$, indicates the component of ${\bf D}$ along 
the direction of the photon propagation and  $\sum_f$ stands for both a sum
over discrete energy levels and an integration over continuum
eigenstates.  For $A > 3$ one faces the
problem that when $\psi_f$ is in the continuum, it cannot be
calculated in an exact way at every excitation energy.  Thus, a direct
calculation of $S(\omega) $ is not possible, unless one introduces
approximations.  However, via the LIT method the problem can be
reformulated in such a way that the knowledge of $\psi_f$ is not
necessary. To this end, one introduces the LIT of the response
function as
\begin{equation} \label{lorenzo}
  {L}(\omega_0,\Gamma )=\int_{\omega_{\rm th}}^{\infty} d\omega \frac{S(\omega)}{(\omega -\omega_0)
               ^2+\Gamma^2}\:\mbox{.}
\end{equation}   
Here $\omega_{\rm th}$ is the threshold energy and $\Gamma > 0 $~is the lorentzian width, 
which plays the role of a resolution parameter. By
using the closure relation one finds
\be
\label{lorenzog} { L}(z)= \langle \psi_0 |
{\hat{D}_0}^{\dagger}\frac{1}{\hat
  {H}-z^*}\frac{1}{\hat{H}-z}\hat{D}_0|\psi_0 \rangle = \langle
\widetilde{\psi} | \widetilde{\psi} \rangle \:\mbox{,} 
\ee
where we introduced the complex energy
$z=E_0+\omega_0+i\Gamma$. The LIT of the response function, $ {L}(z)$,
can be computed directly by solving the Schr{\"o}dinger-like equation
\begin{equation} \label{psi1}
  (\hat{H}-z)|\widetilde{\psi}\rangle =\hat
   {D}_0 | \psi_0\rangle \: \hspace{1cm} 
\end{equation}
for different values of $z$. The solution
$|\widetilde{\psi}\rangle$ has bound-state-like asymptotics,  thus  $L(z)$ can be calculated
even for $A>3$ with any good bound-state technique.  Results for
 $S(\omega)$ are typically obtained from a
numerical inversion~\cite{EfL99,andreasi2005,BarLiv10} of the
LIT~(\ref{lorenzo}) and should be independent
on the choice of $\Gamma$.

We are interested in computing  the dipole response for
$^{16}$O.  To this end, we merge the LIT with the CC
method~\cite{kuemmel1978,bartlett2007,dean2004}, which is a very
efficient bound-state technique, applied with success on several
medium-mass
nuclei~\cite{hagen2008,hagen2010b,hagen2012a,hagen2012b,roth2012}.  In
CC theory the ground state of the system is given by $|\psi_0\ket =
\exp(T)|\phi_0\ket$.  Here $|\phi_0\ket$ is a Slater determinant and
$T$ generates particle-hole ($ph$) excitations. One has $T=T_1+T_2+
\ldots $, as the sum of a $1p$-$1h$ operator $T_1$, a $2p$-$2h$
operator $T_2$, etc.  In what follows we will consider only an
expansion up to $T_2$, which is known as coupled-cluster with
singles-and-doubles (CCSD) excitations~\cite{bartlett2007}.  In CC
theory the LIT of the dipole response function is
\be
\label{cc_lorentz0bar}
  { L}(z)=
\langle 0_L | 
         {\bar{D}_0}^{\dagger}\frac{1}{\bar
              {H}-z^*}\frac{1}{\bar{H} - z}\bar{D}_0 |0_R \rangle
= \langle \widetilde{\psi}_R | \widetilde{\psi}_L \rangle
 \:\mbox{.}
\ee
Here $\bar H = \exp(-T)\hat H\exp(T)$ is the similarity transformed
Hamiltonian, and $\bar D_0^{(\dagger)}=\exp(-T)\hat D^{(\dagger)}_0\exp(T)$ is
the similarity-transformed dipole (dagger) operator.
The states $\langle 0_L | $ and $| 0_R \rangle$ are the left and right
ground-states of the non-hermitian  Hamiltonian
$\bar H$, respectively.  In analogy to Eq.~(\ref{psi1}), we compute
the LIT of the response function by solving a right
Schr{\"o}dinger-like equation
\begin{equation} 
\label{cc_psi1}
  (\bar{H}-z)|\widetilde{\psi}_R(z)\rangle =
  \bar{D}_0 | 0_R\rangle,
\end{equation}
and an equivalent left Schr{\"o}dinger-like equation for
$\langle\widetilde{\psi}_L(z)|$.  Here, $|\widetilde{\psi}_R(z)\ket = R(z)
|\phi_0\ket$, and the operator $R$ is a linear expansion in
particle-hole excitations,
\be R(z) = R_0 + \sum_{i a} R_{i}^{{a}} \hat{c}^\dagger_{{a}}
\hat{c}_{{i}} + {1\over 4}\sum_{i j a b} R_{i j}^{{a b}}
\hat{c}^\dagger_{a}\hat{c}^\dagger_{b} \hat{c}_{j}\hat{c}_{i} + \ldots
\;.  
\ee 
To be consistent with the CCSD approximation we truncate $R(z)$ at the
$2p$-$2h$ excitation level. It is implied that sums over $i,j$ ($a,b$)
run over occupied (unoccupied) states. An equivalent expansion can be
written for the left excitation operator.  The LIT of Eq.~(\ref{cc_lorentz0bar}) can be computed
efficiently by employing a generalization of the Lanczos algorithm for
non-symmetric matrices. 
In our calculations we  use a NN interaction from chiral
effective field theory at next-to-next-to-next-to leading
order (N$^3$LO)~\cite{entem2003}, supplemented by a point Coulomb force. We omit
three-nucleon (3N) forces that already appear at next-to-next-to leading
order~\cite{vankolck1994} due to the problems associated with
computing 3N forces in the large model spaces we employ.
Note that  3N forces have a smaller impact on the 
photo-absorption cross section of $^4$He~\cite{gazit2006} (few percent in
the peak), but it is not clear how large their effect is for $^{16}$O.

\begin{figure}[htb]
\includegraphics[scale=0.5,clip=]{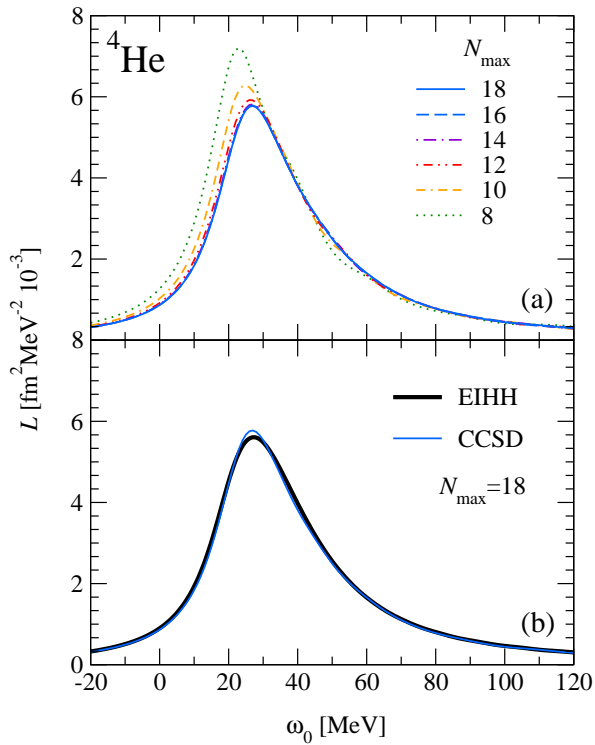}
\caption{(Color online) LIT for  $\Gamma=10$~MeV. (a): Convergence 
 with  $N_{max}$ of the CCSD result with $\hbar \Omega=20$ MeV; (b): Comparison between CCSD and HH results. }
\label{fig_conv}
\end{figure}

{\it Benchmark for $^4$He.}--- For quality assurance, we want to
compare the CC results with virtually exact results obtained from the
effective interaction hyperspherical harmonics (EIHH)
method~\cite{barnea2001} for $^4$He.  Figure~\ref{fig_conv}(a) shows
${L}(\omega_0,\Gamma )$ at fixed
$\Gamma=10$~MeV, computed with the CC method in model spaces with
$\hbar\Omega = 20$~MeV and $N_{\rm max}=2n+l=8, 10, 12, \ldots
18$. The results are well converged in large model spaces with $N_{\rm
  max}=18$.  Figure~\ref{fig_conv}(b) shows a comparison with EIHH results.
The CCSD
calculation is based on a model space with $N_{\rm max}=18$ and
 $\hbar\Omega=20$~MeV and agrees very well with
the EIHH result.  In these large model spaces the
CCSD results are practically independent of the oscillator
frequency.  We attribute the small differences 
to the truncation inherent in the CCSD approximation.

\begin{figure}[htb]
\includegraphics[scale=0.32,clip=]{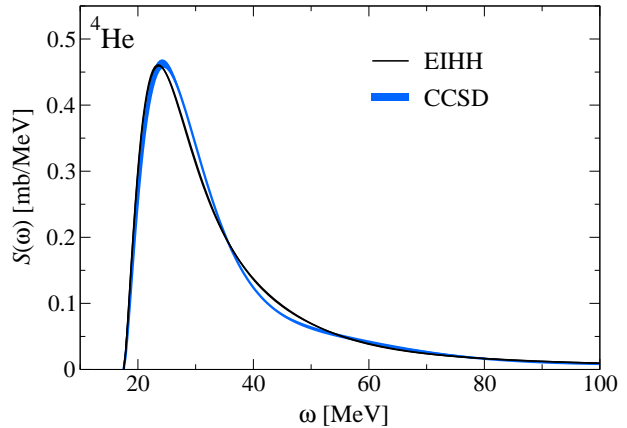}
\caption{(Color online) Comparison of the $^4$He dipole response
  function calculated with CCSD ($\hbar\Omega=20$ MeV and $N_{\rm
    max}=18$) with the EIHH result.  }
\label{fig_comp_HH_resp}
\end{figure}

It is also interesting to compare the two $S(\omega)$ obtained inverting 
the CCSD and HH results for the LIT.  For
the inversion, we use the minimization procedure of
Refs.~\cite{EfL99,andreasi2005}, and require the solutions of this
ill-posed problem to be zero below the threshold energy $\omega_{\rm
  th}$.  The latter is the difference between the binding energy of
$^4$He and $^{3}$H, being $\gamma$ + $^4$He $\rightarrow$ $^3$H + $p$
the first open reaction channel in the inclusive process.  For the
N$^3$LO two-body potential the binding energies are $23.97$ ($7.37$)
MeV obtained within the CCSD approximation and the particle-removed
equation-of-motion method \cite{hagen2010b}, and $25.39$ ($7.85$) MeV
with EIHH for $^4$He ($^3$H), respectively.
Figure~\ref{fig_comp_HH_resp} shows the comparison of the response
functions from the CC and EIHH methods.
The exact threshold energy $\omega_{\rm th}=17.54$ MeV has been
used in all the inversions. For the CCSD calculations we used a model
space with $N_{\rm max}=18$ and $\hbar\Omega=20$ MeV.  We found that
the inversions are insensitive to $\hbar\Omega$.  The band for the
CCSD curve results  from inverting the LIT with
$\Gamma=10$~MeV and $\Gamma=20$~MeV  and estimates the
numerical uncertainty associated with the inversion. For the EIHH
results, the inversions of the LIT with $\Gamma=10$ and 20~MeV
overlap.  The CCSD response function is close to the virtually exact
EIHH result.  Apparently, the small deviations between the CCSD and
the exact result for the LIT in Fig.~\ref{fig_conv}(b) translate into
small deviations in the response function for energies between about
$\omega=30$ and $50$~MeV.

The $^4$He benchmark suggests that the CC method can be employed for
the computation of the dipole response, and that theoretical
uncertainties are well controlled. This makes it exciting to apply the CC method
for the computation of the giant dipole resonance of $^{16}$O.

\begin{figure}[htb]
\includegraphics[scale=0.5,clip=]{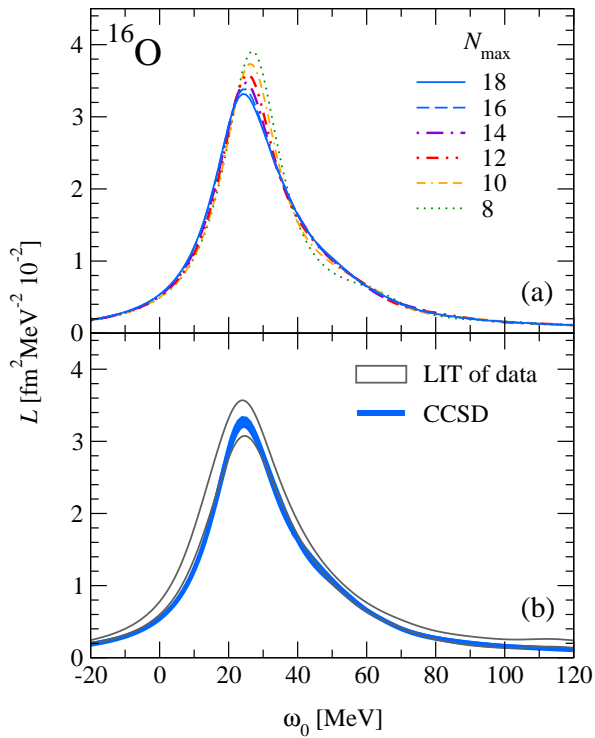}
\caption{(Color online) Convergence of  $L(\omega_0,\Gamma)$ at
  $\Gamma=10$~MeV as a function of  $N_{max}$ for 
  $\hbar \Omega=26$~MeV (a).  Comparison of the LIT at $\Gamma=10$~MeV  within CCSD for $N_{max}=18$  
  and the Lorentz integral transform of Ahrens {\it et al.}~data~\cite{ahrens1975} (b).}
\label{fig_lit_O16}
\end{figure}

{\it Extension to $^{16}\rm{O}$.}---  Figure~\ref{fig_lit_O16}(a) shows the
convergence of the CCSD LIT for $\Gamma=10$~MeV and a sequence of model
spaces with $\hbar\Omega=26$~MeV and $N_{\rm max}$ from 8 up to 18.
The convergence is good, and small differences between $N_{\rm
  max}=16$ and $N_{\rm max}=18$ are only visible close to the maximum.
We recall that the convergence of the LIT also depends on the choice
of the resolution parameter $\Gamma$. A faster convergence is
achieved for increasing $\Gamma$. In what follows, we show
$\Gamma=10$ MeV, the smallest width giving a good convergence.

It is interesting to compare our result of the LIT with data.  The
experiment by Ahrens {\it et al.}~\cite{ahrens1975} measured the total
photo-absorption cross section $\sigma_{\gamma}$ on an Oxygen target
with natural abundance ($99.762\%$ $^{16}$O) with an attenuation
method.  In the unretarded dipole approximation, the cross section is
related to $S(\omega)$ of Eq.~(\ref{response})
via ($\alpha$ is the fine structure constant)
\be 
\sigma_{\gamma}(\omega)=4\pi^2 \alpha \omega S(\omega)\,. 
\label{cs}
\ee 
We extract $S(\omega)$ from the cross-section data, and apply the LIT of
Eq.~(\ref{lorenzo}) to the experimental $S(\omega)$. This allows a direct
comparison with theory, avoiding
the inversion procedure.
Figure~\ref{fig_lit_O16}(b) shows such a comparison.  The band
spanned by the experimental data results from Lorentz-transforming the
data with the corresponding error bars.  The band of the CCSD results
is obtained by varying the harmonic oscillator (HO) frequency between
$\hbar \Omega=20$ and $26$~MeV. 
 The theoretical and experimental
results agree within the uncertainties in almost all the $\omega_0$
range.   
Recall that the (normalized) 
Lorentzian kernel is a representation of the 
$\delta$-function. Therefore the integral in $\omega_0$ of $L(\omega_0,\Gamma)$
is the same as the integral in $\omega$ of $S(\omega)$. Also peak positions are 
approximately conserved.  Thus, from the agreement of the LITs we can already
anticipate that the theoretical results  well reproduce the experimental
centroid and total strength (bremsstrahlung sum rule).

\begin{figure}[htb]
\includegraphics[scale=0.32,clip=]{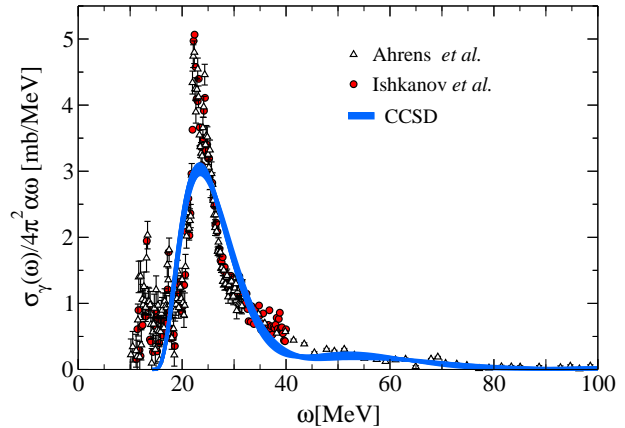}
\caption{(Color online) Comparison of the $^{16}$O dipole
  response calculated in the CCSD scheme  against experimental data by
  Ahrens {\it et al.}~\cite{ahrens1975} (triangles with error bars),
  and Ishkhanov {\it et al.}~\cite{ishkhanov2002} (red circles).}
\label{fig_resp_O16}
\end{figure}

Finally, we perform the inversion of the computed LIT and confront theoretical
and experimental $S(\omega)$.  For the inversion, we make the
following ansatz
\be S(\omega)=\omega^{3/2} \exp \left( -\alpha Z_1Z_2 \sqrt{2\mu\over\omega}\right) 
\sum_i^N c_i e^{-\frac{\omega}{\beta i}}\,. 
\ee
Here, the exponential prefactor is a Gamow factor, and $\mu\sim
\frac{A-1}{A}m_n$ is the reduced mass with $m_n$ being the nucleon
mass. The charges $Z_1=7$ and $Z_2=1$ correspond to the first
disintegration channel (proton separation) in the $\gamma $ $+$
$^{16}$O reaction. The binding energy  and $\omega_{th}$  in the CCSD
approximation (experiment) are 107.24~(127.72)
and $14.25$  (12.13)~MeV, respectively. The threshold energy $\omega_{th}$ is
computed within coupled cluster theory using the particle-removed equation-of-motion~\cite{bartlett2007}.
Note that the inversion is sensitive to numerical noise,
and that the CCSD results using the HO basis are
(only) converged at a few-percent level, and not at an ideal
sub-percent level. In order to improve the convergence and make the
inversion more stable we employ a basis of bound and discretized
continuum states obtained from diagonalizing a spherical Woods-Saxon
potential in a discrete plane-wave basis of 35 mesh points for the
proton and neutron $d_{5/2}, s_{1/2}, d_{3/2}$ partial waves
\cite{hagen2010a}. For the remaining partial waves we employed
$N_{max} =16 $ HO shells with $\hbar\Omega = 26$~MeV.
The inversion determines the coefficients $c_i$ and the non-linear
parameter $\beta$ of the $N$ basis functions by a least-square fit.
Figure~\ref{fig_resp_O16} shows the CCSD dipole response function and
compares it to the data of Ahrens {\it et al.}~\cite{ahrens1975}, and
also to the more recent evaluation by Ishkhanov {\it et
  al.}~\cite{ishkhanov2002,ishkhanov2004}.
The theoretical band is obtained by inverting
the  LIT with 
width  $\Gamma=10$~MeV and by varying the number $N$ of basis
functions employed in the inversion from $5$ to $9$. By inverting the LIT at $\Gamma=20$~MeV we get very similar results. The band is a
lower estimate of the theoretical error, as it does not account e.g.
for missing triples in the cluster expansion.  
The position of the GDR in $^{16}$O is nicely explained by our
calculation whose only ingredient is a NN interaction
which fits NN scattering data. That is the case despite the fact that the binding and threshold 
energies are not correctly reproduced. This fact might be coincidental, given that we are omitting three-nucleon (3N) forces, which may have an impact on the 
response function.
We observe that the form of the theoretical result is somewhat smeared compared to data. 
If larger model spaces were available to allow a more accurate calculation of the LIT at $\Gamma=10$ and even at smaller $\Gamma$, then finer structures in the response could be possibly resolved, if present.
The more structured form of the data below 20 MeV can be due to the contribution of 
higher multipoles (quadrupole and octupole) in the photo-absorption cross section.
We note that
the tail region between
40 and 100~MeV is reproduced within uncertainties.  When integrating
the theoretical photo-absorption cross section up to $100$ MeV we
obtain an enhancement $\kappa=0.57-0.58$ of the Thomas-Reiche-Kuhn sum
rule $\left[59.74 \frac{NZ}{A}{\rm MeV~ mb}(1+\kappa)\right]$.

{\it Summary.}--- We develop a new method based on merging the Lorentz integral transform approach with coupled-cluster theory, 
which opens up important perspectives for studying physical phenomena involving the continuum, also in other fields. 
For example with the present approach one could explore  problems in quantum chemistry, where the CC method is largely used, but the continuum is often discretized. 
 As a first application in nuclear physics
we present a calculation
of the GDR in $^{16}$O,  starting from a NN interaction
from chiral effective field theory that reproduces NN scattering data with 
high accuracy. We compute the Lorentz
integral transform of the dipole response function with the CCSD  scheme. An inversion of the transform yields the
dipole response function. The theoretical result reproduces the
experimental centroid  as well as the total dipole strength
(bremsstrahlung sum rule) very well, and exhibits a somewhat
smeared shape compared to the data. 
This work is a first important step towards an {\it ab-initio} description of the 
electromagnetic responses of medium-heavy stable and rare isotopes from full chiral Hamiltonians.


\begin{acknowledgments}
  We thank Winfried Leidemann for helpful discussions.
  This work was supported in parts by the Natural Sciences
  and Engineering Research Council (NSERC), the National Research
  Council of Canada, the Israel Science Foundation (Grant number
  954/09), the MIUR grant PRIN-2009TWL3MX, the Office of Nuclear
  Physics, U.S.~Department of Energy (Oak Ridge National Laboratory),
  under Grants No.~DE-FG02-96ER40963 (University of Tennessee),
  No.~DE-FC02-07ER41457 (UNEDF SciDAC collaboration), and
  No.~DE-SC0008499 (NUCLEI SciDAC collaboration).  Computer time was
  provided by the Innovative and Novel Computational Impact on Theory
  and Experiment (INCITE) program. This research used resources of the
  Oak Ridge Leadership Computing Facility located in the Oak Ridge
  National Laboratory, which is supported by the Office of Science of
  the Department of Energy under Contract No. DE-AC05-00OR22725, and
  used computational resources of the National Center for
  Computational Sciences, the National Institute for Computational
  Sciences, and TRIUMF.
\end{acknowledgments}

\bibliography{cclit}

\begin{thebibliography}{46}%
\makeatletter
\providecommand \@ifxundefined [1]{%
 \@ifx{#1\undefined}
}%
\providecommand \@ifnum [1]{%
 \ifnum #1\expandafter \@firstoftwo
 \else \expandafter \@secondoftwo
 \fi
}%
\providecommand \@ifx [1]{%
 \ifx #1\expandafter \@firstoftwo
 \else \expandafter \@secondoftwo
 \fi
}%
\providecommand \natexlab [1]{#1}%
\providecommand \enquote  [1]{``#1''}%
\providecommand \bibnamefont  [1]{#1}%
\providecommand \bibfnamefont [1]{#1}%
\providecommand \citenamefont [1]{#1}%
\providecommand \href@noop [0]{\@secondoftwo}%
\providecommand \href [0]{\begingroup \@sanitize@url \@href}%
\providecommand \@href[1]{\@@startlink{#1}\@@href}%
\providecommand \@@href[1]{\endgroup#1\@@endlink}%
\providecommand \@sanitize@url [0]{\catcode `\\12\catcode `\$12\catcode
  `\&12\catcode `\#12\catcode `\^12\catcode `\_12\catcode `\%12\relax}%
\providecommand \@@startlink[1]{}%
\providecommand \@@endlink[0]{}%
\providecommand \url  [0]{\begingroup\@sanitize@url \@url }%
\providecommand \@url [1]{\endgroup\@href {#1}{\urlprefix }}%
\providecommand \urlprefix  [0]{URL }%
\providecommand \Eprint [0]{\href }%
\providecommand \doibase [0]{http://dx.doi.org/}%
\providecommand \selectlanguage [0]{\@gobble}%
\providecommand \bibinfo  [0]{\@secondoftwo}%
\providecommand \bibfield  [0]{\@secondoftwo}%
\providecommand \translation [1]{[#1]}%
\providecommand \BibitemOpen [0]{}%
\providecommand \bibitemStop [0]{}%
\providecommand \bibitemNoStop [0]{.\EOS\space}%
\providecommand \EOS [0]{\spacefactor3000\relax}%
\providecommand \BibitemShut  [1]{\csname bibitem#1\endcsname}%
\let\auto@bib@innerbib\@empty
\bibitem [{\citenamefont {Berman}\ and\ \citenamefont
  {Fultz}(1975)}]{Berman1975}%
  \BibitemOpen
  \bibfield  {author} {\bibinfo {author} {\bibfnamefont {B.~L.}\ \bibnamefont
  {Berman}}\ and\ \bibinfo {author} {\bibfnamefont {S.~C.}\ \bibnamefont
  {Fultz}},\ }\href {\doibase 10.1103/RevModPhys.47.713} {\bibfield  {journal}
  {\bibinfo  {journal} {Rev. Mod. Phys.}\ }\textbf {\bibinfo {volume} {47}},\
  \bibinfo {pages} {713} (\bibinfo {year} {1975})}\BibitemShut {NoStop}%
\bibitem [{\citenamefont {Br{\'e}chignac}\ \emph {et~al.}(1989)\citenamefont
  {Br{\'e}chignac}, \citenamefont {Cahuzac}, \citenamefont {Carlier},\ and\
  \citenamefont {Leygnier}}]{brechignac1989}%
  \BibitemOpen
  \bibfield  {author} {\bibinfo {author} {\bibfnamefont {C.}~\bibnamefont
  {Br{\'e}chignac}}, \bibinfo {author} {\bibfnamefont {P.}~\bibnamefont
  {Cahuzac}}, \bibinfo {author} {\bibfnamefont {F.}~\bibnamefont {Carlier}}, \
  and\ \bibinfo {author} {\bibfnamefont {J.}~\bibnamefont {Leygnier}},\ }\href
  {\doibase 10.1016/0009-2614(89)85233-9} {\bibfield  {journal} {\bibinfo
  {journal} {Chemical Physics Letters}\ }\textbf {\bibinfo {volume} {164}},\
  \bibinfo {pages} {433 } (\bibinfo {year} {1989})}\BibitemShut {NoStop}%
\bibitem [{\citenamefont {Bertsch}\ \emph {et~al.}(1991)\citenamefont
  {Bertsch}, \citenamefont {Bulgac}, \citenamefont {Tom\'anek},\ and\
  \citenamefont {Wang}}]{bertsch1991}%
  \BibitemOpen
  \bibfield  {author} {\bibinfo {author} {\bibfnamefont {G.~F.}\ \bibnamefont
  {Bertsch}}, \bibinfo {author} {\bibfnamefont {A.}~\bibnamefont {Bulgac}},
  \bibinfo {author} {\bibfnamefont {D.}~\bibnamefont {Tom\'anek}}, \ and\
  \bibinfo {author} {\bibfnamefont {Y.}~\bibnamefont {Wang}},\ }\href {\doibase
  10.1103/PhysRevLett.67.2690} {\bibfield  {journal} {\bibinfo  {journal}
  {Phys. Rev. Lett.}\ }\textbf {\bibinfo {volume} {67}},\ \bibinfo {pages}
  {2690} (\bibinfo {year} {1991})}\BibitemShut {NoStop}%
\bibitem [{\citenamefont {Baldwin}\ and\ \citenamefont
  {Klaiber}(1947)}]{BaK47}%
  \BibitemOpen
  \bibfield  {author} {\bibinfo {author} {\bibfnamefont {G.~C.}\ \bibnamefont
  {Baldwin}}\ and\ \bibinfo {author} {\bibfnamefont {G.~S.}\ \bibnamefont
  {Klaiber}},\ }\href {\doibase 10.1103/PhysRev.71.3} {\bibfield  {journal}
  {\bibinfo  {journal} {Phys. Rev.}\ }\textbf {\bibinfo {volume} {71}},\
  \bibinfo {pages} {3} (\bibinfo {year} {1947})}\BibitemShut {NoStop}%
\bibitem [{\citenamefont {Goldhaber}\ and\ \citenamefont
  {Teller}(1948)}]{GoT48}%
  \BibitemOpen
  \bibfield  {author} {\bibinfo {author} {\bibfnamefont {M.}~\bibnamefont
  {Goldhaber}}\ and\ \bibinfo {author} {\bibfnamefont {E.}~\bibnamefont
  {Teller}},\ }\href {\doibase 10.1103/PhysRev.74.1046} {\bibfield  {journal}
  {\bibinfo  {journal} {Phys. Rev.}\ }\textbf {\bibinfo {volume} {74}},\
  \bibinfo {pages} {1046} (\bibinfo {year} {1948})}\BibitemShut {NoStop}%
\bibitem [{\citenamefont {Steinwedel}\ and\ \citenamefont
  {Jensen}(1950)}]{steinwedel1950}%
  \BibitemOpen
  \bibfield  {author} {\bibinfo {author} {\bibfnamefont {H.}~\bibnamefont
  {Steinwedel}}\ and\ \bibinfo {author} {\bibfnamefont {J.~H.~D.}\ \bibnamefont
  {Jensen}},\ }\href@noop {} {\bibfield  {journal} {\bibinfo  {journal} {Z.
  Naturforsch.}\ }\textbf {\bibinfo {volume} {5A}},\ \bibinfo {pages} {413}
  (\bibinfo {year} {1950})}\BibitemShut {NoStop}%
\bibitem [{\citenamefont {Speth}\ and\ \citenamefont {van~der
  Woude}(1981)}]{Speth1981}%
  \BibitemOpen
  \bibfield  {author} {\bibinfo {author} {\bibfnamefont {J.}~\bibnamefont
  {Speth}}\ and\ \bibinfo {author} {\bibfnamefont {A.}~\bibnamefont {van~der
  Woude}},\ }\href {http://stacks.iop.org/0034-4885/44/i=7/a=002} {\bibfield
  {journal} {\bibinfo  {journal} {Reports on Progress in Physics}\ }\textbf
  {\bibinfo {volume} {44}},\ \bibinfo {pages} {719} (\bibinfo {year}
  {1981})}\BibitemShut {NoStop}%
\bibitem [{\citenamefont {Bertsch}\ \emph {et~al.}(1983)\citenamefont
  {Bertsch}, \citenamefont {Bortignon},\ and\ \citenamefont
  {Broglia}}]{Bertsch1983}%
  \BibitemOpen
  \bibfield  {author} {\bibinfo {author} {\bibfnamefont {G.~F.}\ \bibnamefont
  {Bertsch}}, \bibinfo {author} {\bibfnamefont {P.~F.}\ \bibnamefont
  {Bortignon}}, \ and\ \bibinfo {author} {\bibfnamefont {R.~A.}\ \bibnamefont
  {Broglia}},\ }\href {\doibase 10.1103/RevModPhys.55.287} {\bibfield
  {journal} {\bibinfo  {journal} {Rev. Mod. Phys.}\ }\textbf {\bibinfo {volume}
  {55}},\ \bibinfo {pages} {287} (\bibinfo {year} {1983})}\BibitemShut
  {NoStop}%
\bibitem [{\citenamefont {Erler}\ \emph {et~al.}(2011)\citenamefont {Erler},
  \citenamefont {Kl\"{u}pfel},\ and\ \citenamefont {Reinhard}}]{Erler2011}%
  \BibitemOpen
  \bibfield  {author} {\bibinfo {author} {\bibfnamefont {J.}~\bibnamefont
  {Erler}}, \bibinfo {author} {\bibfnamefont {P.}~\bibnamefont {Kl\"{u}pfel}},
  \ and\ \bibinfo {author} {\bibfnamefont {P.-G.}\ \bibnamefont {Reinhard}},\
  }\href {\doibase doi:10.1088/0954-3899/38/3/033101} {\bibfield  {journal}
  {\bibinfo  {journal} {J. Phys. G: Nucl. Part. Phys.}\ }\textbf {\bibinfo
  {volume} {38}},\ \bibinfo {pages} {033101} (\bibinfo {year} {2011})},\
  \bibinfo {note} {and references therein}\BibitemShut {NoStop}%
\bibitem [{\citenamefont {Nakatsukasa}(2012)}]{Nakatsukasa2012}%
  \BibitemOpen
  \bibfield  {author} {\bibinfo {author} {\bibfnamefont {T.}~\bibnamefont
  {Nakatsukasa}},\ }\href {\doibase 10.1093/ptep/pts016} {\bibfield  {journal}
  {\bibinfo  {journal} {Prog. Theor. Exp. Phys.}\ }\textbf {\bibinfo {volume}
  {01A207}} (\bibinfo {year} {2012}),\ 10.1093/ptep/pts016},\ \bibinfo {note}
  {and references therein}\BibitemShut {NoStop}%
\bibitem [{\citenamefont {Shlomo}\ and\ \citenamefont
  {Bertsch}(1975)}]{Shlomo}%
  \BibitemOpen
  \bibfield  {author} {\bibinfo {author} {\bibfnamefont {S.}~\bibnamefont
  {Shlomo}}\ and\ \bibinfo {author} {\bibfnamefont {G.}~\bibnamefont
  {Bertsch}},\ }\href@noop {} {\bibfield  {journal} {\bibinfo  {journal} {Nucl.
  Phys.}\ }\textbf {\bibinfo {volume} {A243}},\ \bibinfo {pages} {507}
  (\bibinfo {year} {1975})}\BibitemShut {NoStop}%
\bibitem [{\citenamefont {Col\'o}\ and\ \citenamefont
  {Bortignon}(2001)}]{Colo2001}%
  \BibitemOpen
  \bibfield  {author} {\bibinfo {author} {\bibfnamefont {G.}~\bibnamefont
  {Col\'o}}\ and\ \bibinfo {author} {\bibfnamefont {P.~F.}\ \bibnamefont
  {Bortignon}},\ }\href@noop {} {\bibfield  {journal} {\bibinfo  {journal}
  {Nucl. Phys.}\ }\textbf {\bibinfo {volume} {A696}},\ \bibinfo {pages} {427}
  (\bibinfo {year} {2001})}\BibitemShut {NoStop}%
\bibitem [{\citenamefont {Lyutorovich}\ \emph {et~al.}(2012)\citenamefont
  {Lyutorovich}, \citenamefont {Tselyaev}, \citenamefont {Speth}, \citenamefont
  {Krewald}, \citenamefont {Gr\"ummer},\ and\ \citenamefont
  {Reinhard}}]{LyT12}%
  \BibitemOpen
  \bibfield  {author} {\bibinfo {author} {\bibfnamefont {N.}~\bibnamefont
  {Lyutorovich}}, \bibinfo {author} {\bibfnamefont {V.~I.}\ \bibnamefont
  {Tselyaev}}, \bibinfo {author} {\bibfnamefont {J.}~\bibnamefont {Speth}},
  \bibinfo {author} {\bibfnamefont {S.}~\bibnamefont {Krewald}}, \bibinfo
  {author} {\bibfnamefont {F.}~\bibnamefont {Gr\"ummer}}, \ and\ \bibinfo
  {author} {\bibfnamefont {P.-G.}\ \bibnamefont {Reinhard}},\ }\href {\doibase
  10.1103/PhysRevLett.109.092502} {\bibfield  {journal} {\bibinfo  {journal}
  {Phys. Rev. Lett.}\ }\textbf {\bibinfo {volume} {109}},\ \bibinfo {pages}
  {092502} (\bibinfo {year} {2012})}\BibitemShut {NoStop}%
\bibitem [{\citenamefont {Efros}\ \emph {et~al.}(1994)\citenamefont {Efros},
  \citenamefont {Leidemann},\ and\ \citenamefont {Orlandini}}]{efros1994}%
  \BibitemOpen
  \bibfield  {author} {\bibinfo {author} {\bibfnamefont {V.~D.}\ \bibnamefont
  {Efros}}, \bibinfo {author} {\bibfnamefont {W.}~\bibnamefont {Leidemann}}, \
  and\ \bibinfo {author} {\bibfnamefont {G.}~\bibnamefont {Orlandini}},\ }\href
  {\doibase 10.1016/0370-2693(94)91355-2} {\bibfield  {journal} {\bibinfo
  {journal} {Physics Letters B}\ }\textbf {\bibinfo {volume} {338}},\ \bibinfo
  {pages} {130 } (\bibinfo {year} {1994})}\BibitemShut {NoStop}%
\bibitem [{\citenamefont {Efros}\ \emph {et~al.}(2007)\citenamefont {Efros},
  \citenamefont {Leidemann}, \citenamefont {Orlandini},\ and\ \citenamefont
  {Barnea}}]{Efl07}%
  \BibitemOpen
  \bibfield  {author} {\bibinfo {author} {\bibfnamefont {V.~D.}\ \bibnamefont
  {Efros}}, \bibinfo {author} {\bibfnamefont {W.}~\bibnamefont {Leidemann}},
  \bibinfo {author} {\bibfnamefont {G.}~\bibnamefont {Orlandini}}, \ and\
  \bibinfo {author} {\bibfnamefont {N.}~\bibnamefont {Barnea}},\ }\href
  {http://stacks.iop.org/0954-3899/34/i=12/a=R02} {\bibfield  {journal}
  {\bibinfo  {journal} {Journal of Physics G: Nuclear and Particle Physics}\
  }\textbf {\bibinfo {volume} {34}},\ \bibinfo {pages} {R459} (\bibinfo {year}
  {2007})}\BibitemShut {NoStop}%
\bibitem [{\citenamefont {Barnea}\ \emph {et~al.}(2001)\citenamefont {Barnea},
  \citenamefont {Leidemann},\ and\ \citenamefont {Orlandini}}]{barnea2001}%
  \BibitemOpen
  \bibfield  {author} {\bibinfo {author} {\bibfnamefont {N.}~\bibnamefont
  {Barnea}}, \bibinfo {author} {\bibfnamefont {W.}~\bibnamefont {Leidemann}}, \
  and\ \bibinfo {author} {\bibfnamefont {G.}~\bibnamefont {Orlandini}},\ }\href
  {\doibase 10.1016/S0375-9474(01)00794-1} {\bibfield  {journal} {\bibinfo
  {journal} {Nuclear Physics A}\ }\textbf {\bibinfo {volume} {693}},\ \bibinfo
  {pages} {565 } (\bibinfo {year} {2001})}\BibitemShut {NoStop}%
\bibitem [{\citenamefont {Fenin}\ and\ \citenamefont
  {Efros}(1972)}]{fenin1972}%
  \BibitemOpen
  \bibfield  {author} {\bibinfo {author} {\bibfnamefont {Y.~I.}\ \bibnamefont
  {Fenin}}\ and\ \bibinfo {author} {\bibfnamefont {V.~D.}\ \bibnamefont
  {Efros}},\ }\href@noop {} {\bibfield  {journal} {\bibinfo  {journal} {Sov. J.
  Nucl. Phys.}\ }\textbf {\bibinfo {volume} {15}},\ \bibinfo {pages} {497}
  (\bibinfo {year} {1972})}\BibitemShut {NoStop}%
\bibitem [{\citenamefont {Navr\'atil}\ and\ \citenamefont
  {Battett}(1996)}]{ncsm}%
  \BibitemOpen
  \bibfield  {author} {\bibinfo {author} {\bibfnamefont {P.}~\bibnamefont
  {Navr\'atil}}\ and\ \bibinfo {author} {\bibfnamefont {B.~R.}\ \bibnamefont
  {Battett}},\ }\href {\doibase 10.1103/PhysRevC.54.2986} {\bibfield  {journal}
  {\bibinfo  {journal} {Phys. Rev. C}\ }\textbf {\bibinfo {volume} {54}},\
  \bibinfo {pages} {2986} (\bibinfo {year} {1996})}\BibitemShut {NoStop}%
\bibitem [{\citenamefont {Efros}\ \emph {et~al.}(1997)\citenamefont {Efros},
  \citenamefont {Leidemann},\ and\ \citenamefont {Orlandini}}]{EfL97}%
  \BibitemOpen
  \bibfield  {author} {\bibinfo {author} {\bibfnamefont {V.~D.}\ \bibnamefont
  {Efros}}, \bibinfo {author} {\bibfnamefont {W.}~\bibnamefont {Leidemann}}, \
  and\ \bibinfo {author} {\bibfnamefont {G.}~\bibnamefont {Orlandini}},\ }\href
  {\doibase 10.1103/PhysRevLett.78.4015} {\bibfield  {journal} {\bibinfo
  {journal} {Phys. Rev. Lett.}\ }\textbf {\bibinfo {volume} {78}},\ \bibinfo
  {pages} {4015} (\bibinfo {year} {1997})}\BibitemShut {NoStop}%
\bibitem [{\citenamefont {Stetcu}\ \emph {et~al.}(2007)\citenamefont {Stetcu},
  \citenamefont {Quaglioni}, \citenamefont {Bacca}, \citenamefont {Barrett},
  \citenamefont {Johnson}, \citenamefont {Navr{\'a}til}, \citenamefont
  {Barnea}, \citenamefont {Leidemann},\ and\ \citenamefont
  {Orlandini}}]{stetcu2007}%
  \BibitemOpen
  \bibfield  {author} {\bibinfo {author} {\bibfnamefont {I.}~\bibnamefont
  {Stetcu}}, \bibinfo {author} {\bibfnamefont {S.}~\bibnamefont {Quaglioni}},
  \bibinfo {author} {\bibfnamefont {S.}~\bibnamefont {Bacca}}, \bibinfo
  {author} {\bibfnamefont {B.~R.}\ \bibnamefont {Barrett}}, \bibinfo {author}
  {\bibfnamefont {C.~W.}\ \bibnamefont {Johnson}}, \bibinfo {author}
  {\bibfnamefont {P.}~\bibnamefont {Navr{\'a}til}}, \bibinfo {author}
  {\bibfnamefont {N.}~\bibnamefont {Barnea}}, \bibinfo {author} {\bibfnamefont
  {W.}~\bibnamefont {Leidemann}}, \ and\ \bibinfo {author} {\bibfnamefont
  {G.}~\bibnamefont {Orlandini}},\ }\href {\doibase
  10.1016/j.nuclphysa.2006.12.047} {\bibfield  {journal} {\bibinfo  {journal}
  {Nuclear Physics A}\ }\textbf {\bibinfo {volume} {785}},\ \bibinfo {pages}
  {307 } (\bibinfo {year} {2007})}\BibitemShut {NoStop}%
\bibitem [{\citenamefont {Bacca}\ \emph {et~al.}(2002)\citenamefont {Bacca},
  \citenamefont {Marchisio}, \citenamefont {Barnea}, \citenamefont
  {Leidemann},\ and\ \citenamefont {Orlandini}}]{bacca2002}%
  \BibitemOpen
  \bibfield  {author} {\bibinfo {author} {\bibfnamefont {S.}~\bibnamefont
  {Bacca}}, \bibinfo {author} {\bibfnamefont {M.~A.}\ \bibnamefont
  {Marchisio}}, \bibinfo {author} {\bibfnamefont {N.}~\bibnamefont {Barnea}},
  \bibinfo {author} {\bibfnamefont {W.}~\bibnamefont {Leidemann}}, \ and\
  \bibinfo {author} {\bibfnamefont {G.}~\bibnamefont {Orlandini}},\ }\href
  {\doibase 10.1103/PhysRevLett.89.052502} {\bibfield  {journal} {\bibinfo
  {journal} {Phys. Rev. Lett.}\ }\textbf {\bibinfo {volume} {89}},\ \bibinfo
  {pages} {052502} (\bibinfo {year} {2002})}\BibitemShut {NoStop}%
\bibitem [{\citenamefont {Bacca}\ \emph
  {et~al.}(2004{\natexlab{a}})\citenamefont {Bacca}, \citenamefont {Barnea},
  \citenamefont {Leidemann},\ and\ \citenamefont {Orlandini}}]{BaB04}%
  \BibitemOpen
  \bibfield  {author} {\bibinfo {author} {\bibfnamefont {S.}~\bibnamefont
  {Bacca}}, \bibinfo {author} {\bibfnamefont {N.}~\bibnamefont {Barnea}},
  \bibinfo {author} {\bibfnamefont {W.}~\bibnamefont {Leidemann}}, \ and\
  \bibinfo {author} {\bibfnamefont {G.}~\bibnamefont {Orlandini}},\ }\href
  {\doibase 10.1103/PhysRevC.69.057001} {\bibfield  {journal} {\bibinfo
  {journal} {Phys. Rev. C}\ }\textbf {\bibinfo {volume} {69}},\ \bibinfo
  {pages} {057001} (\bibinfo {year} {2004}{\natexlab{a}})}\BibitemShut
  {NoStop}%
\bibitem [{\citenamefont {Bacca}\ \emph
  {et~al.}(2004{\natexlab{b}})\citenamefont {Bacca}, \citenamefont
  {Arenh{\"o}vel}, \citenamefont {Barnea}, \citenamefont {Leidemann},\ and\
  \citenamefont {Orlandini}}]{BaA04}%
  \BibitemOpen
  \bibfield  {author} {\bibinfo {author} {\bibfnamefont {S.}~\bibnamefont
  {Bacca}}, \bibinfo {author} {\bibfnamefont {H.}~\bibnamefont
  {Arenh{\"o}vel}}, \bibinfo {author} {\bibfnamefont {N.}~\bibnamefont
  {Barnea}}, \bibinfo {author} {\bibfnamefont {W.}~\bibnamefont {Leidemann}}, \
  and\ \bibinfo {author} {\bibfnamefont {G.}~\bibnamefont {Orlandini}},\ }\href
  {\doibase 10.1016/j.physletb.2004.10.025} {\bibfield  {journal} {\bibinfo
  {journal} {Physics Letters B}\ }\textbf {\bibinfo {volume} {603}},\ \bibinfo
  {pages} {159 } (\bibinfo {year} {2004}{\natexlab{b}})}\BibitemShut {NoStop}%
\bibitem [{\citenamefont {Gazit}\ \emph {et~al.}(2006)\citenamefont {Gazit},
  \citenamefont {Bacca}, \citenamefont {Barnea}, \citenamefont {Leidemann},\
  and\ \citenamefont {Orlandini}}]{gazit2006}%
  \BibitemOpen
  \bibfield  {author} {\bibinfo {author} {\bibfnamefont {D.}~\bibnamefont
  {Gazit}}, \bibinfo {author} {\bibfnamefont {S.}~\bibnamefont {Bacca}},
  \bibinfo {author} {\bibfnamefont {N.}~\bibnamefont {Barnea}}, \bibinfo
  {author} {\bibfnamefont {W.}~\bibnamefont {Leidemann}}, \ and\ \bibinfo
  {author} {\bibfnamefont {G.}~\bibnamefont {Orlandini}},\ }\href {\doibase
  10.1103/PhysRevLett.96.112301} {\bibfield  {journal} {\bibinfo  {journal}
  {Phys. Rev. Lett.}\ }\textbf {\bibinfo {volume} {96}},\ \bibinfo {pages}
  {112301} (\bibinfo {year} {2006})}\BibitemShut {NoStop}%
\bibitem [{\citenamefont {Quaglioni}\ and\ \citenamefont
  {Navr\'atil}(2007)}]{Qua07}%
  \BibitemOpen
  \bibfield  {author} {\bibinfo {author} {\bibfnamefont {S.}~\bibnamefont
  {Quaglioni}}\ and\ \bibinfo {author} {\bibfnamefont {P.}~\bibnamefont
  {Navr\'atil}},\ }\href {\doibase 10.1016/j.physletb.2007.06.082} {\bibfield
  {journal} {\bibinfo  {journal} {Phys. Lett. B}\ }\textbf {\bibinfo {volume}
  {652}},\ \bibinfo {pages} {370} (\bibinfo {year} {2007})}\BibitemShut
  {NoStop}%
\bibitem [{\citenamefont {Sagawa}\ and\ \citenamefont
  {Suzuki}(1999)}]{sagawa1999}%
  \BibitemOpen
  \bibfield  {author} {\bibinfo {author} {\bibfnamefont {H.}~\bibnamefont
  {Sagawa}}\ and\ \bibinfo {author} {\bibfnamefont {T.}~\bibnamefont
  {Suzuki}},\ }\href {\doibase 10.1103/PhysRevC.59.3116} {\bibfield  {journal}
  {\bibinfo  {journal} {Phys. Rev. C}\ }\textbf {\bibinfo {volume} {59}},\
  \bibinfo {pages} {3116} (\bibinfo {year} {1999})}\BibitemShut {NoStop}%
\bibitem [{\citenamefont {Leistenschneider}\ \emph {et~al.}(2001)\citenamefont
  {Leistenschneider}, \citenamefont {Aumann}, \citenamefont {Boretzky},
  \citenamefont {Cortina}, \citenamefont {Cub}, \citenamefont {Pramanik},
  \citenamefont {Dostal}, \citenamefont {Elze}, \citenamefont {Emling},
  \citenamefont {Geissel}, \citenamefont {Gr\"unschlo\ss{}}, \citenamefont
  {Hellstr}, \citenamefont {Holzmann}, \citenamefont {Ilievski}, \citenamefont
  {Iwasa}, \citenamefont {Kaspar}, \citenamefont {Kleinb\"ohl}, \citenamefont
  {Kratz}, \citenamefont {Kulessa}, \citenamefont {Leifels}, \citenamefont
  {Lubkiewicz}, \citenamefont {M\"unzenberg}, \citenamefont {Reiter},
  \citenamefont {Rejmund}, \citenamefont {Scheidenberger}, \citenamefont
  {Schlegel}, \citenamefont {Simon}, \citenamefont {Stroth}, \citenamefont
  {S\"ummerer}, \citenamefont {Wajda}, \citenamefont {Wal\'us},\ and\
  \citenamefont {Wan}}]{Lei01}%
  \BibitemOpen
  \bibfield  {author} {\bibinfo {author} {\bibfnamefont {A.}~\bibnamefont
  {Leistenschneider}}, \bibinfo {author} {\bibfnamefont {T.}~\bibnamefont
  {Aumann}}, \bibinfo {author} {\bibfnamefont {K.}~\bibnamefont {Boretzky}},
  \bibinfo {author} {\bibfnamefont {D.}~\bibnamefont {Cortina}}, \bibinfo
  {author} {\bibfnamefont {J.}~\bibnamefont {Cub}}, \bibinfo {author}
  {\bibfnamefont {U.~D.}\ \bibnamefont {Pramanik}}, \bibinfo {author}
  {\bibfnamefont {W.}~\bibnamefont {Dostal}}, \bibinfo {author} {\bibfnamefont
  {T.~W.}\ \bibnamefont {Elze}}, \bibinfo {author} {\bibfnamefont
  {H.}~\bibnamefont {Emling}}, \bibinfo {author} {\bibfnamefont
  {H.}~\bibnamefont {Geissel}}, \bibinfo {author} {\bibfnamefont
  {A.}~\bibnamefont {Gr\"unschlo\ss{}}}, \bibinfo {author} {\bibfnamefont
  {M.}~\bibnamefont {Hellstr}}, \bibinfo {author} {\bibfnamefont
  {R.}~\bibnamefont {Holzmann}}, \bibinfo {author} {\bibfnamefont
  {S.}~\bibnamefont {Ilievski}}, \bibinfo {author} {\bibfnamefont
  {N.}~\bibnamefont {Iwasa}}, \bibinfo {author} {\bibfnamefont
  {M.}~\bibnamefont {Kaspar}}, \bibinfo {author} {\bibfnamefont
  {A.}~\bibnamefont {Kleinb\"ohl}}, \bibinfo {author} {\bibfnamefont {J.~V.}\
  \bibnamefont {Kratz}}, \bibinfo {author} {\bibfnamefont {R.}~\bibnamefont
  {Kulessa}}, \bibinfo {author} {\bibfnamefont {Y.}~\bibnamefont {Leifels}},
  \bibinfo {author} {\bibfnamefont {E.}~\bibnamefont {Lubkiewicz}}, \bibinfo
  {author} {\bibfnamefont {G.}~\bibnamefont {M\"unzenberg}}, \bibinfo {author}
  {\bibfnamefont {P.}~\bibnamefont {Reiter}}, \bibinfo {author} {\bibfnamefont
  {M.}~\bibnamefont {Rejmund}}, \bibinfo {author} {\bibfnamefont
  {C.}~\bibnamefont {Scheidenberger}}, \bibinfo {author} {\bibfnamefont
  {C.}~\bibnamefont {Schlegel}}, \bibinfo {author} {\bibfnamefont
  {H.}~\bibnamefont {Simon}}, \bibinfo {author} {\bibfnamefont
  {J.}~\bibnamefont {Stroth}}, \bibinfo {author} {\bibfnamefont
  {K.}~\bibnamefont {S\"ummerer}}, \bibinfo {author} {\bibfnamefont
  {E.}~\bibnamefont {Wajda}}, \bibinfo {author} {\bibfnamefont
  {W.}~\bibnamefont {Wal\'us}}, \ and\ \bibinfo {author} {\bibfnamefont
  {S.}~\bibnamefont {Wan}},\ }\href {\doibase 10.1103/PhysRevLett.86.5442}
  {\bibfield  {journal} {\bibinfo  {journal} {Phys. Rev. Lett.}\ }\textbf
  {\bibinfo {volume} {86}},\ \bibinfo {pages} {5442} (\bibinfo {year}
  {2001})}\BibitemShut {NoStop}%
\bibitem [{\citenamefont {K{\"u}mmel}\ \emph {et~al.}(1978)\citenamefont
  {K{\"u}mmel}, \citenamefont {L{\"u}hrmann},\ and\ \citenamefont
  {Zabolitzky}}]{kuemmel1978}%
  \BibitemOpen
  \bibfield  {author} {\bibinfo {author} {\bibfnamefont {H.}~\bibnamefont
  {K{\"u}mmel}}, \bibinfo {author} {\bibfnamefont {K.~H.}\ \bibnamefont
  {L{\"u}hrmann}}, \ and\ \bibinfo {author} {\bibfnamefont {J.~G.}\
  \bibnamefont {Zabolitzky}},\ }\href {\doibase 10.1016/0370-1573(78)90081-9}
  {\bibfield  {journal} {\bibinfo  {journal} {Physics Reports}\ }\textbf
  {\bibinfo {volume} {36}},\ \bibinfo {pages} {1 } (\bibinfo {year}
  {1978})}\BibitemShut {NoStop}%
\bibitem [{\citenamefont {Bartlett}\ and\ \citenamefont
  {Musia\l{}}(2007)}]{bartlett2007}%
  \BibitemOpen
  \bibfield  {author} {\bibinfo {author} {\bibfnamefont {R.~J.}\ \bibnamefont
  {Bartlett}}\ and\ \bibinfo {author} {\bibfnamefont {M.}~\bibnamefont
  {Musia\l{}}},\ }\href {\doibase 10.1103/RevModPhys.79.291} {\bibfield
  {journal} {\bibinfo  {journal} {Rev. Mod. Phys.}\ }\textbf {\bibinfo {volume}
  {79}},\ \bibinfo {pages} {291} (\bibinfo {year} {2007})}\BibitemShut
  {NoStop}%
\bibitem [{\citenamefont {Epelbaum}\ \emph {et~al.}(2009)\citenamefont
  {Epelbaum}, \citenamefont {Hammer},\ and\ \citenamefont
  {Mei\ss{}ner}}]{epelbaum2009}%
  \BibitemOpen
  \bibfield  {author} {\bibinfo {author} {\bibfnamefont {E.}~\bibnamefont
  {Epelbaum}}, \bibinfo {author} {\bibfnamefont {H.-W.}\ \bibnamefont
  {Hammer}}, \ and\ \bibinfo {author} {\bibfnamefont {U.-G.}\ \bibnamefont
  {Mei\ss{}ner}},\ }\href {\doibase 10.1103/RevModPhys.81.1773} {\bibfield
  {journal} {\bibinfo  {journal} {Rev. Mod. Phys.}\ }\textbf {\bibinfo {volume}
  {81}},\ \bibinfo {pages} {1773} (\bibinfo {year} {2009})}\BibitemShut
  {NoStop}%
\bibitem [{\citenamefont {Machleidt}\ and\ \citenamefont
  {Entem}(2011)}]{machleidt2011}%
  \BibitemOpen
  \bibfield  {author} {\bibinfo {author} {\bibfnamefont {R.}~\bibnamefont
  {Machleidt}}\ and\ \bibinfo {author} {\bibfnamefont {D.}~\bibnamefont
  {Entem}},\ }\href {\doibase 10.1016/j.physrep.2011.02.001} {\bibfield
  {journal} {\bibinfo  {journal} {Physics Reports}\ }\textbf {\bibinfo {volume}
  {503}},\ \bibinfo {pages} {1 } (\bibinfo {year} {2011})}\BibitemShut
  {NoStop}%
\bibitem [{\citenamefont {Efros}\ \emph {et~al.}(1999)\citenamefont {Efros},
  \citenamefont {Leidemann},\ and\ \citenamefont {Orlandini}}]{EfL99}%
  \BibitemOpen
  \bibfield  {author} {\bibinfo {author} {\bibfnamefont {V.~D.}\ \bibnamefont
  {Efros}}, \bibinfo {author} {\bibfnamefont {W.}~\bibnamefont {Leidemann}}, \
  and\ \bibinfo {author} {\bibfnamefont {G.}~\bibnamefont {Orlandini}},\ }\href
  {\doibase 10.1007/s006010050118} {\bibfield  {journal} {\bibinfo  {journal}
  {Few-Body Systems}\ }\textbf {\bibinfo {volume} {26}},\ \bibinfo {pages}
  {251} (\bibinfo {year} {1999})}\BibitemShut {NoStop}%
\bibitem [{\citenamefont {Andreasi}\ \emph {et~al.}(2005)\citenamefont
  {Andreasi}, \citenamefont {Leidemann}, \citenamefont {Rei{\ss}},\ and\
  \citenamefont {Schwamb}}]{andreasi2005}%
  \BibitemOpen
  \bibfield  {author} {\bibinfo {author} {\bibfnamefont {D.}~\bibnamefont
  {Andreasi}}, \bibinfo {author} {\bibfnamefont {W.}~\bibnamefont {Leidemann}},
  \bibinfo {author} {\bibfnamefont {C.}~\bibnamefont {Rei{\ss}}}, \ and\
  \bibinfo {author} {\bibfnamefont {M.}~\bibnamefont {Schwamb}},\ }\href
  {\doibase 10.1140/epja/i2005-10009-3} {\bibfield  {journal} {\bibinfo
  {journal} {The European Physical Journal A - Hadrons and Nuclei}\ }\textbf
  {\bibinfo {volume} {24}},\ \bibinfo {pages} {361} (\bibinfo {year}
  {2005})}\BibitemShut {NoStop}%
\bibitem [{\citenamefont {Barnea}\ and\ \citenamefont
  {Liverts}(2010)}]{BarLiv10}%
  \BibitemOpen
  \bibfield  {author} {\bibinfo {author} {\bibfnamefont {N.}~\bibnamefont
  {Barnea}}\ and\ \bibinfo {author} {\bibfnamefont {E.}~\bibnamefont
  {Liverts}},\ }\href {\doibase 10.1007/s00601-010-0090-z} {\bibfield
  {journal} {\bibinfo  {journal} {Few-Body Systems}\ }\textbf {\bibinfo
  {volume} {48}},\ \bibinfo {pages} {11} (\bibinfo {year} {2010})}\BibitemShut
  {NoStop}%
\bibitem [{\citenamefont {Dean}\ and\ \citenamefont
  {Hjorth-Jensen}(2004)}]{dean2004}%
  \BibitemOpen
  \bibfield  {author} {\bibinfo {author} {\bibfnamefont {D.~J.}\ \bibnamefont
  {Dean}}\ and\ \bibinfo {author} {\bibfnamefont {M.}~\bibnamefont
  {Hjorth-Jensen}},\ }\href {\doibase 10.1103/PhysRevC.69.054320} {\bibfield
  {journal} {\bibinfo  {journal} {Phys. Rev. C}\ }\textbf {\bibinfo {volume}
  {69}},\ \bibinfo {pages} {054320} (\bibinfo {year} {2004})}\BibitemShut
  {NoStop}%
\bibitem [{\citenamefont {Hagen}\ \emph {et~al.}(2008)\citenamefont {Hagen},
  \citenamefont {Papenbrock}, \citenamefont {Dean},\ and\ \citenamefont
  {Hjorth-Jensen}}]{hagen2008}%
  \BibitemOpen
  \bibfield  {author} {\bibinfo {author} {\bibfnamefont {G.}~\bibnamefont
  {Hagen}}, \bibinfo {author} {\bibfnamefont {T.}~\bibnamefont {Papenbrock}},
  \bibinfo {author} {\bibfnamefont {D.~J.}\ \bibnamefont {Dean}}, \ and\
  \bibinfo {author} {\bibfnamefont {M.}~\bibnamefont {Hjorth-Jensen}},\ }\href
  {\doibase 10.1103/PhysRevLett.101.092502} {\bibfield  {journal} {\bibinfo
  {journal} {Phys. Rev. Lett.}\ }\textbf {\bibinfo {volume} {101}},\ \bibinfo
  {pages} {092502} (\bibinfo {year} {2008})}\BibitemShut {NoStop}%
\bibitem [{\citenamefont {Hagen}\ \emph
  {et~al.}(2010{\natexlab{a}})\citenamefont {Hagen}, \citenamefont
  {Papenbrock}, \citenamefont {Dean},\ and\ \citenamefont
  {Hjorth-Jensen}}]{hagen2010b}%
  \BibitemOpen
  \bibfield  {author} {\bibinfo {author} {\bibfnamefont {G.}~\bibnamefont
  {Hagen}}, \bibinfo {author} {\bibfnamefont {T.}~\bibnamefont {Papenbrock}},
  \bibinfo {author} {\bibfnamefont {D.~J.}\ \bibnamefont {Dean}}, \ and\
  \bibinfo {author} {\bibfnamefont {M.}~\bibnamefont {Hjorth-Jensen}},\ }\href
  {\doibase 10.1103/PhysRevC.82.034330} {\bibfield  {journal} {\bibinfo
  {journal} {Phys. Rev. C}\ }\textbf {\bibinfo {volume} {82}},\ \bibinfo
  {pages} {034330} (\bibinfo {year} {2010}{\natexlab{a}})}\BibitemShut
  {NoStop}%
\bibitem [{\citenamefont {Hagen}\ \emph
  {et~al.}(2012{\natexlab{a}})\citenamefont {Hagen}, \citenamefont
  {Hjorth-Jensen}, \citenamefont {Jansen}, \citenamefont {Machleidt},\ and\
  \citenamefont {Papenbrock}}]{hagen2012a}%
  \BibitemOpen
  \bibfield  {author} {\bibinfo {author} {\bibfnamefont {G.}~\bibnamefont
  {Hagen}}, \bibinfo {author} {\bibfnamefont {M.}~\bibnamefont
  {Hjorth-Jensen}}, \bibinfo {author} {\bibfnamefont {G.~R.}\ \bibnamefont
  {Jansen}}, \bibinfo {author} {\bibfnamefont {R.}~\bibnamefont {Machleidt}}, \
  and\ \bibinfo {author} {\bibfnamefont {T.}~\bibnamefont {Papenbrock}},\
  }\href {\doibase 10.1103/PhysRevLett.108.242501} {\bibfield  {journal}
  {\bibinfo  {journal} {Phys. Rev. Lett.}\ }\textbf {\bibinfo {volume} {108}},\
  \bibinfo {pages} {242501} (\bibinfo {year} {2012}{\natexlab{a}})}\BibitemShut
  {NoStop}%
\bibitem [{\citenamefont {Hagen}\ \emph
  {et~al.}(2012{\natexlab{b}})\citenamefont {Hagen}, \citenamefont
  {Hjorth-Jensen}, \citenamefont {Jansen}, \citenamefont {Machleidt},\ and\
  \citenamefont {Papenbrock}}]{hagen2012b}%
  \BibitemOpen
  \bibfield  {author} {\bibinfo {author} {\bibfnamefont {G.}~\bibnamefont
  {Hagen}}, \bibinfo {author} {\bibfnamefont {M.}~\bibnamefont
  {Hjorth-Jensen}}, \bibinfo {author} {\bibfnamefont {G.~R.}\ \bibnamefont
  {Jansen}}, \bibinfo {author} {\bibfnamefont {R.}~\bibnamefont {Machleidt}}, \
  and\ \bibinfo {author} {\bibfnamefont {T.}~\bibnamefont {Papenbrock}},\
  }\href {\doibase 10.1103/PhysRevLett.109.032502} {\bibfield  {journal}
  {\bibinfo  {journal} {Phys. Rev. Lett.}\ }\textbf {\bibinfo {volume} {109}},\
  \bibinfo {pages} {032502} (\bibinfo {year} {2012}{\natexlab{b}})}\BibitemShut
  {NoStop}%
\bibitem [{\citenamefont {Roth}\ \emph {et~al.}(2012)\citenamefont {Roth},
  \citenamefont {Binder}, \citenamefont {Vobig}, \citenamefont {Calci},
  \citenamefont {Langhammer},\ and\ \citenamefont {Navr\'atil}}]{roth2012}%
  \BibitemOpen
  \bibfield  {author} {\bibinfo {author} {\bibfnamefont {R.}~\bibnamefont
  {Roth}}, \bibinfo {author} {\bibfnamefont {S.}~\bibnamefont {Binder}},
  \bibinfo {author} {\bibfnamefont {K.}~\bibnamefont {Vobig}}, \bibinfo
  {author} {\bibfnamefont {A.}~\bibnamefont {Calci}}, \bibinfo {author}
  {\bibfnamefont {J.}~\bibnamefont {Langhammer}}, \ and\ \bibinfo {author}
  {\bibfnamefont {P.}~\bibnamefont {Navr\'atil}},\ }\href {\doibase
  10.1103/PhysRevLett.109.052501} {\bibfield  {journal} {\bibinfo  {journal}
  {Phys. Rev. Lett.}\ }\textbf {\bibinfo {volume} {109}},\ \bibinfo {pages}
  {052501} (\bibinfo {year} {2012})}\BibitemShut {NoStop}%
\bibitem [{\citenamefont {Entem}\ and\ \citenamefont
  {Machleidt}(2003)}]{entem2003}%
  \BibitemOpen
  \bibfield  {author} {\bibinfo {author} {\bibfnamefont {D.~R.}\ \bibnamefont
  {Entem}}\ and\ \bibinfo {author} {\bibfnamefont {R.}~\bibnamefont
  {Machleidt}},\ }\href {\doibase 10.1103/PhysRevC.68.041001} {\bibfield
  {journal} {\bibinfo  {journal} {Phys. Rev. C}\ }\textbf {\bibinfo {volume}
  {68}},\ \bibinfo {pages} {041001} (\bibinfo {year} {2003})}\BibitemShut
  {NoStop}%
\bibitem [{\citenamefont {van Kolck}(1994)}]{vankolck1994}%
  \BibitemOpen
  \bibfield  {author} {\bibinfo {author} {\bibfnamefont {U.}~\bibnamefont {van
  Kolck}},\ }\href {\doibase 10.1103/PhysRevC.49.2932} {\bibfield  {journal}
  {\bibinfo  {journal} {Phys. Rev. C}\ }\textbf {\bibinfo {volume} {49}},\
  \bibinfo {pages} {2932} (\bibinfo {year} {1994})}\BibitemShut {NoStop}%
\bibitem [{\citenamefont {Ahrens}\ \emph {et~al.}(1975)\citenamefont {Ahrens},
  \citenamefont {Borchert}, \citenamefont {Czock}, \citenamefont {Eppler},
  \citenamefont {Gimm}, \citenamefont {Gundrum}, \citenamefont {Kr{\"o}ning},
  \citenamefont {Riehn}, \citenamefont {Ram}, \citenamefont {Zieger},\ and\
  \citenamefont {Ziegler}}]{ahrens1975}%
  \BibitemOpen
  \bibfield  {author} {\bibinfo {author} {\bibfnamefont {J.}~\bibnamefont
  {Ahrens}}, \bibinfo {author} {\bibfnamefont {H.}~\bibnamefont {Borchert}},
  \bibinfo {author} {\bibfnamefont {K.}~\bibnamefont {Czock}}, \bibinfo
  {author} {\bibfnamefont {H.}~\bibnamefont {Eppler}}, \bibinfo {author}
  {\bibfnamefont {H.}~\bibnamefont {Gimm}}, \bibinfo {author} {\bibfnamefont
  {H.}~\bibnamefont {Gundrum}}, \bibinfo {author} {\bibfnamefont
  {M.}~\bibnamefont {Kr{\"o}ning}}, \bibinfo {author} {\bibfnamefont
  {P.}~\bibnamefont {Riehn}}, \bibinfo {author} {\bibfnamefont {G.~S.}\
  \bibnamefont {Ram}}, \bibinfo {author} {\bibfnamefont {A.}~\bibnamefont
  {Zieger}}, \ and\ \bibinfo {author} {\bibfnamefont {B.}~\bibnamefont
  {Ziegler}},\ }\href {\doibase 10.1016/0375-9474(75)90543-6} {\bibfield
  {journal} {\bibinfo  {journal} {Nuclear Physics A}\ }\textbf {\bibinfo
  {volume} {251}},\ \bibinfo {pages} {479 } (\bibinfo {year}
  {1975})}\BibitemShut {NoStop}%
\bibitem [{\citenamefont {Ishkhanov}\ \emph {et~al.}(2002)\citenamefont
  {Ishkhanov}, \citenamefont {Kapitonov}, \citenamefont {Lileeva},
  \citenamefont {Shirokov}, \citenamefont {Erokhova}, \citenamefont {Elkin},\
  and\ \citenamefont {Izotova}}]{ishkhanov2002}%
  \BibitemOpen
  \bibfield  {author} {\bibinfo {author} {\bibfnamefont {B.~S.}\ \bibnamefont
  {Ishkhanov}}, \bibinfo {author} {\bibfnamefont {I.~M.}\ \bibnamefont
  {Kapitonov}}, \bibinfo {author} {\bibfnamefont {E.~I.}\ \bibnamefont
  {Lileeva}}, \bibinfo {author} {\bibfnamefont {E.~V.}\ \bibnamefont
  {Shirokov}}, \bibinfo {author} {\bibfnamefont {V.~A.}\ \bibnamefont
  {Erokhova}}, \bibinfo {author} {\bibfnamefont {M.~A.}\ \bibnamefont {Elkin}},
  \ and\ \bibinfo {author} {\bibfnamefont {A.~V.}\ \bibnamefont {Izotova}},\
  }\href@noop {} {\emph {\bibinfo {title} {Cross sections of photon absorption
  by nuclei with nucleon numbers 12 - 65}}},\ \bibinfo {type} {Tech. Rep.}\
  \bibinfo {number} {MSU-INP-2002-27/711}\ (\bibinfo  {institution} {Institute
  of Nuclear Physics},\ \bibinfo {address} {Moscow State University},\ \bibinfo
  {year} {2002})\BibitemShut {NoStop}%
\bibitem [{\citenamefont {Hagen}\ \emph
  {et~al.}(2010{\natexlab{b}})\citenamefont {Hagen}, \citenamefont
  {Papenbrock},\ and\ \citenamefont {Hjorth-Jensen}}]{hagen2010a}%
  \BibitemOpen
  \bibfield  {author} {\bibinfo {author} {\bibfnamefont {G.}~\bibnamefont
  {Hagen}}, \bibinfo {author} {\bibfnamefont {T.}~\bibnamefont {Papenbrock}}, \
  and\ \bibinfo {author} {\bibfnamefont {M.}~\bibnamefont {Hjorth-Jensen}},\
  }\href {\doibase 10.1103/PhysRevLett.104.182501} {\bibfield  {journal}
  {\bibinfo  {journal} {Phys. Rev. Lett.}\ }\textbf {\bibinfo {volume} {104}},\
  \bibinfo {pages} {182501} (\bibinfo {year} {2010}{\natexlab{b}})}\BibitemShut
  {NoStop}%
\bibitem [{\citenamefont {Ishkhanov}\ and\ \citenamefont
  {Orlin}(2004)}]{ishkhanov2004}%
  \BibitemOpen
  \bibfield  {author} {\bibinfo {author} {\bibfnamefont {B.}~\bibnamefont
  {Ishkhanov}}\ and\ \bibinfo {author} {\bibfnamefont {V.}~\bibnamefont
  {Orlin}},\ }\href {\doibase 10.1134/1.1755384} {\bibfield  {journal}
  {\bibinfo  {journal} {Physics of Atomic Nuclei}\ }\textbf {\bibinfo {volume}
  {67}},\ \bibinfo {pages} {920} (\bibinfo {year} {2004})}\BibitemShut
  {NoStop}%
\end{thebibliography}%
\bibliographystyle{apsrev4-1}

\end{document}